# Sub-barrier cavitation in liquid helium

**Mikhail Pekker and Mikhail N. Shneider[1*],**

[1]*Department of Mechanical and Aerospace Engineering, Princeton University, Princeton, NJ USA*
[*] m.n.shneider@gmail.com

## Abstract

In this paper, the tunneling mechanism of cavitation in liquid helium for $^3$He and $^4$He is considered on the basis of the Schrödinger-like equation. It is assumed that the pairwise interactions of helium atoms are determined by the Lennard-Jones potential. The kinetics of nucleation and the mechanism that limits the growth of cavitation bubbles in liquid helium are considered, taking into account their growth in a negative pressure field.

## Introduction

The classical description of nucleation, or the formation of growing bubbles in a liquid under negative pressure, faces an insurmountable contradiction at low temperatures. From the classical point of view, the number of bubbles with a critical radius generated per unit volume per unit time is proportional to $e^{-W_{cr}/k_BT}$, where $W_{cr}=16\pi\sigma/3P_-^2$ is the minimum activation energy required to create a growing bubble; $k_B$ is the Boltzmann constant; $T$ is the temperature of the liquid; $\sigma$ is the surface tension coefficient; and $P_-=P_{in}-P_{out}$ is the negative pressure, which is equal to the pressure difference between the inside and outside of the bubble [1]. Since $W_{cr}$ is practically independent of the fluid's temperature, a greater negative pressure is required to form bubbles of critical size as the temperature decreases. However, experiments [2] have shown that at a certain fixed temperature, the nucleation rate in liquid helium ceases to depend on temperature.

A similar occurrence takes place in chemistry. In classical chemistry, the rate of a chemical reaction depends exponentially on the ratio of the activation energy $I_{act}$ and the temperature $T$ of the reagents: $C\propto e^{-I_{act}/k_BT}$ (the Arrhenius law). According to the Arrhenius law, the reaction rate should decrease as the temperature of the reagents decreases until it ceases completely. Experiments indeed have shown an exponential decrease in the rate of chemical reactions with a decrease in temperature. However, starting from a certain temperature, the reaction rate ceases to fall and remains constant [3,4].

These two facts can be explained based on quantum mechanics. Chemical reactions at low temperatures, for which the Arrhenius law is not applicable, are possible due to tunneling, as Hund first noted in 1927 [4]. Since then, the effect of sub-barrier tunneling in "cold chemistry" has been discussed in numerous theoretical and experimental articles (see reviews [3,4]). Lifshits and Kagan first considered the possibility of tunnel nucleation (creation of growing bubbles) in liquid helium [6], and this concept has been further developed in many articles (see, for example, [7–11]).

In the original paper by Lifshitz and Kagan [6] and subsequent papers [9, 11], liquid helium was considered within the framework of a continuous medium with a constant surface tension coefficient independent of pore size and tensile negative pressure in the quasi-classical approximation. In [6] and subsequent works, the bubble was considered as a particle of variable mass, $M$ in a potential, $U$, as shown in Figure 1. In the quasi-classical approximation, the energies $E_0$, $E_1$, and $E_2$ in Figure 1 are determined from the equation:

$$n+\frac{1}{2}=\frac{1}{\pi\hbar}\int_0^{R_{1.n}}\sqrt{2M(E_n-U)}dr, \qquad n=0,1,2,\dots . \tag{1}$$

Accordingly, the probability of a sub-barrier transition from an energy state corresponding to $E_n$ to a bubble is proportional to

$$W_n \sim exp\left(-\frac{2}{\hbar}\int_{R_{1,n}}^{R_{2,n}}\sqrt{2M(U-E_n)}dr\right). \quad (2)$$

In equations (1) and (2), ħ is Planck's constant. The $R_{1,n}$ values represent the points at which the particle reflects off the potential barrier. The $R_{1,n}$ and $R_{2,n}$ values represent the coordinates of the sub-barrier transition of the particle with energy $E_n$. Expressions (1) and (2) allow only a qualitative assessment of nucleation at temperatures close to zero, since the quasi-classical approximation is valid only at $n \gg 1$. Expressions for $M$ and $U$ will be presented in the following sections of the paper.

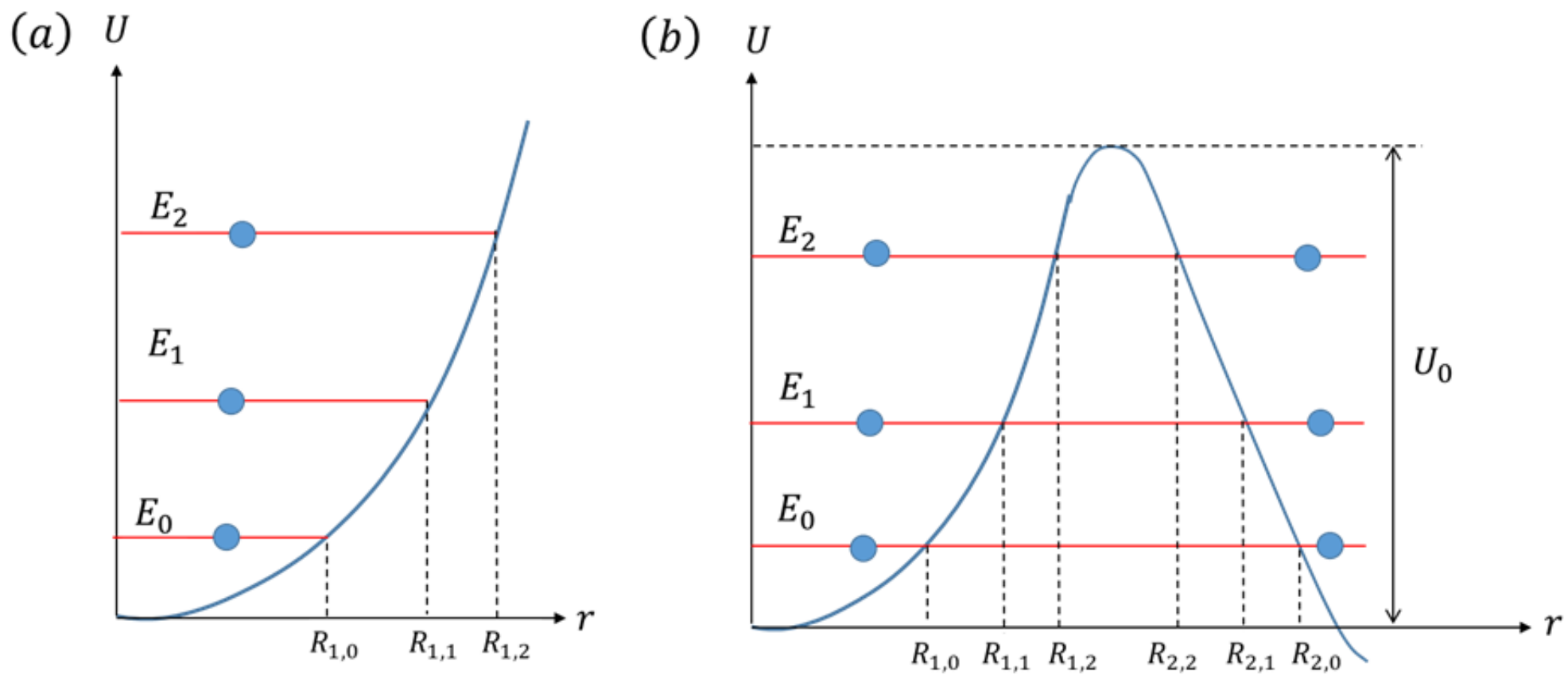

**Figure 1.** A particle in a potential well. (a) The potential well is infinitely deep. A particle can only be in a potential well. (b) The potential well with a potential barrier. The particle can be both inside and outside the well. $U_0$ is the height of the potential barrier. The energy $E_0$ corresponds to the particle's vibration energy at the zero energy level. The energies $E_1$ and $E_2$ correspond to the first and second excited levels, respectively. $R_{1,0}$, $R_{1,1}$, and $R_{2,2}$ are the reflection points of the particle in the potential well. The segments $[R_{1,0}\ R_{2,0}]$, $[R_{1,1}\ R_{2,1}]$, and $[R_{1,2}\ R_{2,2}]$ on (b) correspond to sub-barrier transitions from levels $E_0$, $E_1$, and $E_2$, respectively.

The probability of a particle being at the first excited level, as shown in Fig. 1, is $\sim e^{-(E_1-E_0)/k_BT}$. The probability of a particle being at the second level is $\sim e^{-(E_2-E_0)/k_BT}$. The following relations are valid for cryogenic liquids, such as $^4$He and $^3$He:

$$\frac{E_1-E_0}{k_BT} \gg 1\ , \frac{E_2-E_0}{k_BT} >>1. \quad (3)$$

Therefore, although tunneling through the barrier for a particle located at levels 1 and 2 is more likely than for a particle at level 0, the total probability of a particle passing through level 0 at low temperatures is much greater. For ordinary liquids, such as water, the freezing point is much higher than the energy difference between the excited and lower (ground) levels. Thus, the quantum properties of liquids at nucleation are weakly expressed.

Note that the presence of a superfluid phase in $^4$He does not affect cavitation bubble formation. This is because the surface tension coefficient of bubbles has no peculiarities and varies by only 10% within the 0–2 K region [12].

In this paper, we propose a wave equation that describes the tunneling mechanism of cavitation in liquid helium. This equation is analogous to the stationary Schrödinger equation for a point particle. The eigenvalues and wave functions are found for various values of negative pressure in $^4$He and $^3$He. We develop the Zel'dovich-Fischer theory of nucleation for liquid helium, considering the effect of tunneling through the potential barrier. Additionally, we examine the mechanism that restricts bubble formation and consider bubble expansion in the negative pressure field.

## I. Surface tension of a bubble in a negative pressure field

In the field of negative pressure, the liquid is stretched, and the value of $n_w = \rho_{|P_-|}/m$ changes, where $\rho_{|P_-|}$ is the density of the liquid in the field of negative pressure $P_-$ and $m$ is the mass of the atoms or molecules constituting the liquid. The relationship between the densities of $^4$He and $^3$He and negative pressure $P_-$ can be obtained by integrating the equations for the speed of sound $c$ [2].

$$\left(\frac{\partial P}{\partial \rho}\right)_T = c^2_{He^4} = \left(14.3(P + 9.5 \cdot 10^5)\right)^{2/3} \text{ for } ^4\text{He}, \quad (4)$$

and

$$\left(\frac{\partial P}{\partial \rho}\right)_T = c^2_{He^3} = \left(19.23(P + 3 \cdot 10^5)\right)^{2/3} \text{ for } ^3\text{He}. \quad (5)$$

In formulas (4) and (5), pressure is in Pascals, and the sound speeds and the speeds of sound $c_{He^4}$ and $c_{He^3}$ are in meters per second. Using equations (4) and (5) and counting the pressure from atmospheric pressure, we can derive the following equations of state for $^4$He and $^3$He, respectively:

$$P_- = 7.57(\rho_{|P_-|} - 93.24)^3 - 9.5 \cdot 10^5 - p_{atm} = 7.57(\rho_{|P_-|} - 93.24)^3 - 10.5 \cdot 10^5 \quad ^4\text{He and} \quad (6)$$

$$P_- = 13.7(\rho_{|P_-|} - 51.55)^3 - 3.0 \cdot 10^5 - p_{atm} = 13.7(\rho_{|P_-|} - 51.55)^3 - 4 \cdot 10^5 \quad ^3\text{He}. \quad (7)$$

The dependencies of $^4$He and $^3$He densities on $|P_-|$ are shown in Fig. 2(a) and Fig. 2(b), respectively.

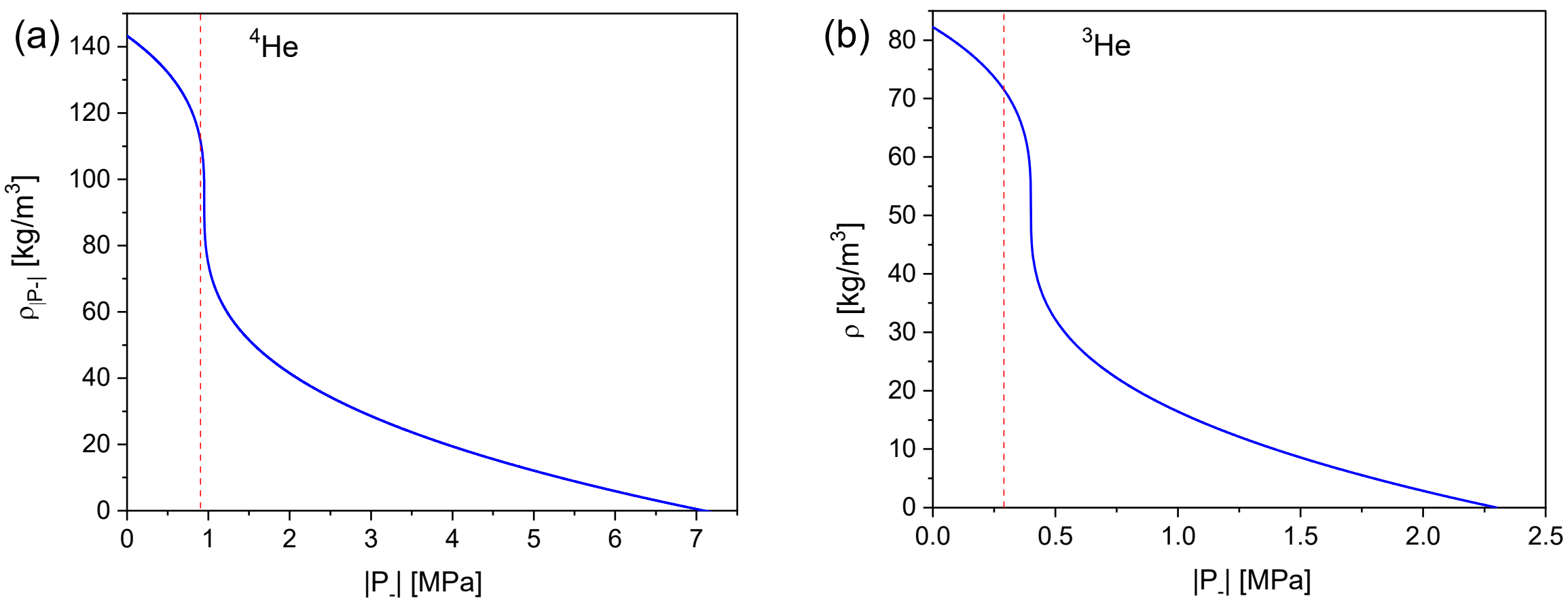

**Figure 2.** Density dependencies for liquid helium: (a) for $^4$He and (b) for $^3$He. The vertical red dashed lines indicate the maximum absolute values of negative pressure at which cavitation begins [2].

In [14,15], the surface tension coefficients for a flat boundary and for a bubble of finite radius were calculated under the assumption of the Lennard-Jones potential, which determines the pairwise interaction between atoms [16]:

$$W = 4W_m\left(\left(\frac{r_0}{r}\right)^{12} - \left(\frac{r_0}{r}\right)^{6}\right), \tag{8}$$

for helium in the gas phase: $W_m = 14.1 \cdot 10^{-23}$J and $r_0 = 0.287 \cdot 10^{-9}$m [17]. As shown in references [18,19], at a temperature close to zero for liquid $^4$He, $r_0$ is practically the same as in the gaseous phase. The corresponding values for $W_m$ are $8.75 \cdot 10^{-23}$J [17] and $7.54 \cdot 10^{-23}$ J [19], are close to the experimental value of $W_m = 9.17 \cdot 10^{-23}$ J [20].

Figure 3(a) and (b) show the dependence of the surface tension coefficient, $\sigma_{0,|P_-|}$, on the flat boundary for $^4$He and $^3$He, respectively.

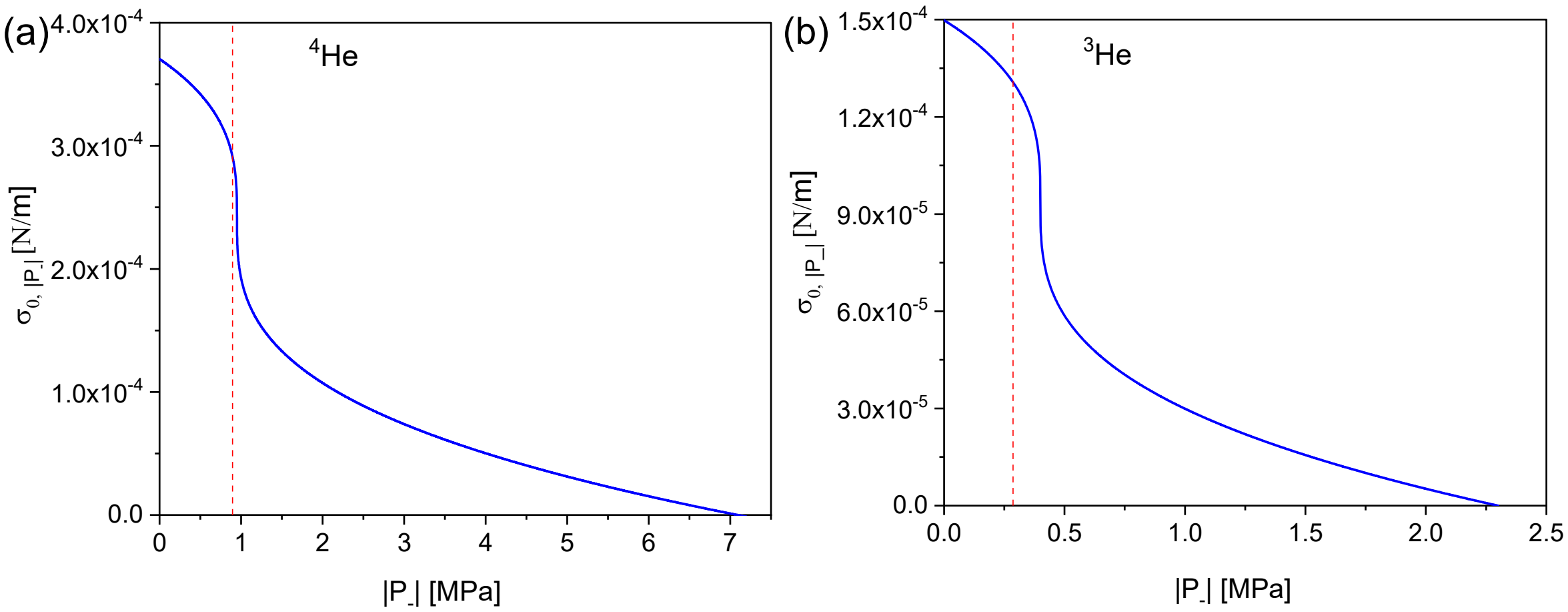

**Figure 3.** Surface tension dependencies for a flat interface for liquid helium are shown in panels (a) for $^4$He and (b) for $^3$He. The vertical red dashed lines indicate the maximum absolute values of negative pressure at which cavitation begins [2].

Figure 4 shows the dimensionless dependence of the surface tension coefficient, $\xi_b$, on the bubble radius related to $\sigma_{0,|P_-|}$. This dependence is universal to all liquids, the interaction of whose molecules is determined by the Lennard-Jones potential (see references [14,15]).

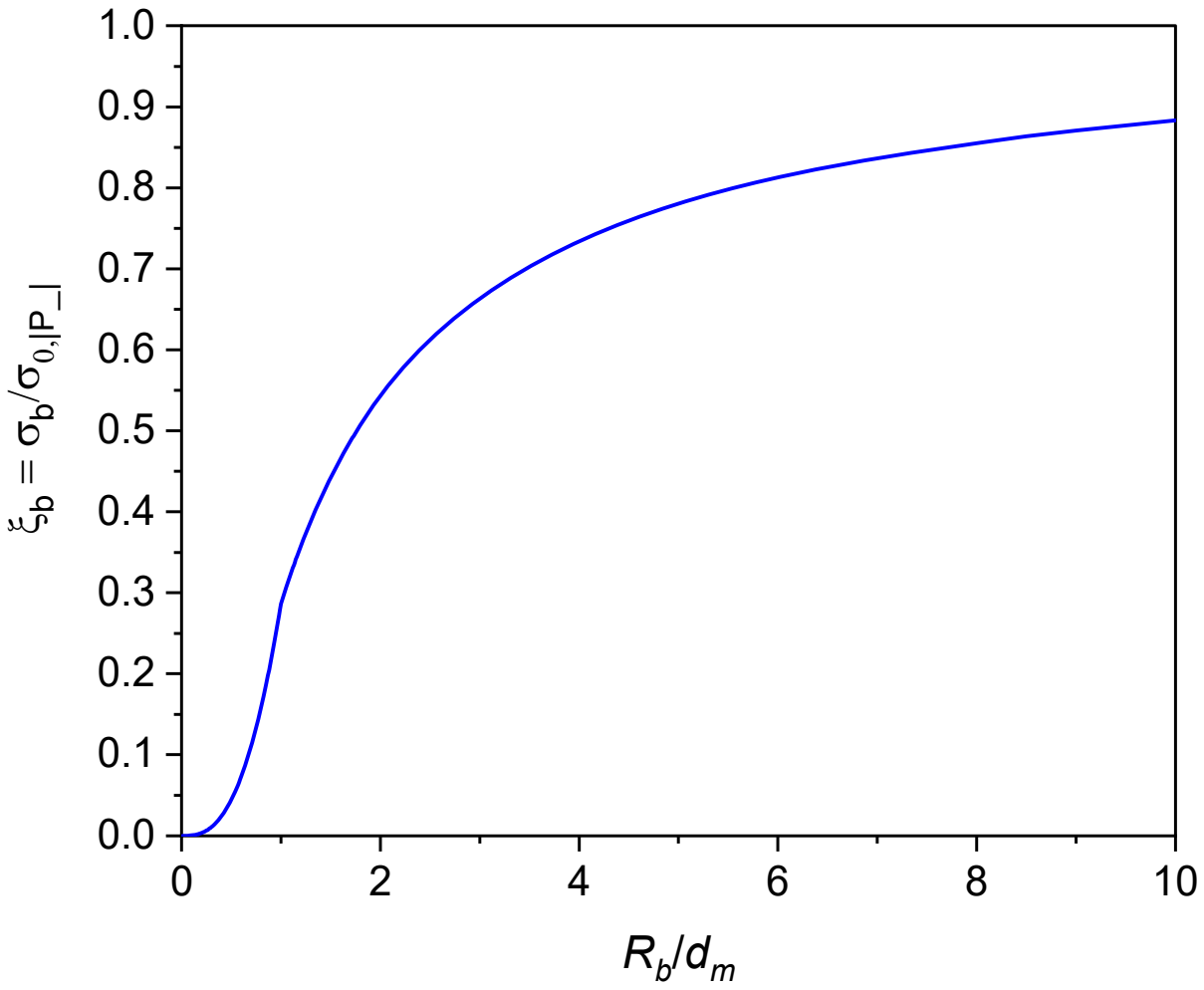

**Figure 4.** The dependences of the surface tension coefficient $\sigma_b$ normalized to the surface tension coefficient of a flat interface $\sigma_{0,|P_-|}$ on the bubble radius $R_b/d_m$, $d_m = 2^{1/6}r_0$.

## II. Statement of the problem

As in [6,9,11], below, we will consider the liquid to be continuous and the cavitation bubbles to be voids with infinitely thin boundaries. The added mass will be concentrated in these boundaries, and the surface tension coefficient will depend on the bubble radius and negative pressure.

Follow [6] consider the Rayleigh bubble. The kinetic energy associated with the movement of the added mass is [8,13-16]:

$$E_{kin} = 2\pi R_b^3 \rho \left(\frac{dR_b}{dt}\right)^2, \tag{9}$$

where $R_b$ is the radius of the bubble and $\rho$ is the density of the liquid.

Assuming that the interaction between the atoms is determined by the Lennard-Jones potential, the potential energy of the cavitation pores is given by:

$$W_b = 8\pi\sigma_{|P_-|} \int_{d_m}^{R_b} r\xi_b(r)\, dr - \frac{4\pi}{3} |P_-| (R_b - d_m)^3, \tag{10}$$

where $|P_-|$ is the absolute value of the negative pressure and $\sigma_{|P_-|}$ is the surface tension coefficient of the liquid in the negative pressure field $P_-$. In equation (10), we account for the fact that the radius of the bubble, $R_b$, cannot be smaller than the effective diameter of the molecule, $d_m = 2^{1/6} r_0$. For a constant surface tension coefficient, $\sigma_0$, independent of pore size, negative pressure, and the effective diameter of the molecule ($d_m = 0$), formula (10) reduces to the following form:

$$W_{0,b} = 4\pi\sigma_0 R_b^2 - \frac{4\pi}{3} |P_-| R_b^3 . \tag{11}$$

In accordance with [6], the Lagrange function for a pore can be formally written as [14-16]:

$$L(R_b, \dot{R}_b) = E_{kin} - W_b = \frac{4\pi R_b^3 \rho_{|P_-|} \dot{R}_b^2}{2} - 8\pi\sigma_{|P_-|} \int_{d_m}^{R_b} r\xi_b(r)\, dr + \frac{4\pi}{3} |P_-| (R_b - d_m)^3. \tag{12}$$

The expression $4\pi R_b^3 \rho_{|P_-|}$ in equation (12) can be considered an attached mass concentrated within a thin shell of radius $R_b$ that moves with a velocity $v = \dot{R}_b$. The mass of the shell, $M = 4\pi R_b^3 \rho_{|P_-|}$, depends on its radius.

From (12), we can derive the canonical momentum $p_b$:

$$p_b = -\frac{\partial L}{\partial \dot{R}_b} = -2\pi \rho_{|P_-|} R_b^3 \dot{R}_b. \tag{13}$$

The corresponding Hamiltonian to the Lagrangian (12) is:

$$H(R_b, P_b) = \frac{p_b^2}{2M} + W_b = \frac{4\pi \rho_{|P_-|} R_b^3}{2} \dot{R}_b^2 + 8\pi\sigma_{|P_-|} \int_{d_m}^{R_b} r\xi_b(r)\, dr - \frac{4\pi}{3} |P_-| (R_b - d_m)^3. \tag{14}$$

When the surface tension coefficient $\sigma_0$ and liquid density $\rho_0$ are constant and independent of $R_b$ and $P_-$, the Hamiltonian (14) coincides with the one introduced in [6]:

$$H_0(R_b, P_b) = \frac{p_b^2}{2M} + W_{0,b} = \frac{4\pi \rho_0 R_b^3}{2} \dot{R}_b^2 + 4\pi\sigma_0 R_b^2 - \frac{4\pi}{3} |P_-| R_b^3. \tag{15}$$

In the original work of Lifshits and Kagan [6] and subsequent works [9, 11], equation (15) was considered within the context of Bohr's approach to calculating the energy of zero-point oscillations, as described by formulas (1) and (2).

The Hamiltonian (14) can be considered the Hamiltonian of a particle of mass $M$, which depends on the coordinate $R_b$, located in a given potential $W_b$. Following the analogy of quantum mechanics, when the Hamiltonian (14) corresponds to the stationary Schrödinger equation for a point particle, we can assume that equation (14) corresponds to the following equation [14-16]:

$$\frac{1}{R_b^2}\frac{d}{dR_b}R_b^2\frac{d\Psi}{dR_b} = -\left(\frac{2M}{\hbar^2}(E - W_b)\right)\Psi =$$

$$-\left(\frac{8\pi\rho_{|P_-|}R_b^3}{\hbar^2}\left(E - 8\pi\sigma_{|P_-|}\int_{d_m}^{R_b} r\xi_b(r)\,dr + \frac{4\pi}{3}|P_-|(R_b - d_m)^3\right)\right)\Psi, \tag{16}$$

where $\Psi$ is the wave function. Equation (16) differs from the standard Schrödinger equation in that the particle is considered point-like, and the square of the wave function's modulus describes the probability of finding the particle at a given point in space. In our case, which is described by equation (16), the "particles" correspond to a hollow sphere with an infinitely thin shell of the mass $M$. The square of the modulus of the wave function describes the probability that the shell will have a radius of $R_b$. If the mass $M$ is constant, then equation (16) becomes the usual Schrödinger equation.

In the Lennard-Jones model, the function $\xi_b$ is universal; that is, it depends only on $R_b/d_m$. Therefore, it is convenient to introduce the variable $x = R_b/d_m$. Writing the wave function as $\Psi = \chi/R_b$, we obtain:

$$\frac{d^2\chi}{dx^2} + \alpha x^3\left(E' - \frac{\sigma_{|P_-|}}{\sigma_0}\int_1^x \xi_b F(\xi)\,d\xi + \beta(x-1)^3\right)\chi = 0, \tag{17}$$

where $\alpha = \frac{64\pi^2 d_m^7\rho_{|P_-|}\sigma_0}{\hbar^2}$, $\beta = \frac{d_m}{6\sigma_0}|P_-|$, $E' = \frac{E}{8\pi\sigma_0 d_m^2}$.

Eigenvalues of the energy and temperature dependencies of the critical negative pressure, $P_{cr}$, corresponding to equation (17) for $^4He$ and $^3He$, are presented in [14,15].

## III. Eigenfunctions

Equation (17), like the Schrödinger equation, have eigenvalues that determine the possible energy levels of a particle/thin shell in a potential well. The eigenvalues of energy in degrees Kelvin corresponding to equation (17) can be found in references [14,15].

Figures 5(a) and 6(a) show the potential: $U_b = \frac{W_b}{k_B} = \frac{\alpha}{k_B}\left(\frac{\sigma_{|P_-|}}{\sigma_0}\int_1^x \xi F(\xi)\,d\xi - \beta(x-1)^3\right)$ in Kelvin degrees, corresponding to equation (17), and eigenfunction modes for $^4He$ and $^3He$, respectively. Unlike the Coulomb potential, only a finite number of modes always exist in this potential. The deeper the potential well, the smaller the $|P_-|$, and the more modes with higher energy exist in it. Furthermore, each new mode arises with an energy equal to the height of the potential barrier and "goes down" with respect to the height of the barrier as the barrier grows, as shown in Figure 7. Figures 5(b)–5(d) and 6(b)–6(d) show the wave functions that correspond to the eigenmodes depicted in Figures 5(a) and 6(a). The horizontal lines in Figures 5(a) and 6(a) show the eigenenergies corresponding to the first mode for $P_- = 1.05$ MPa, 1.04 MPa, and 1.02 MPa.

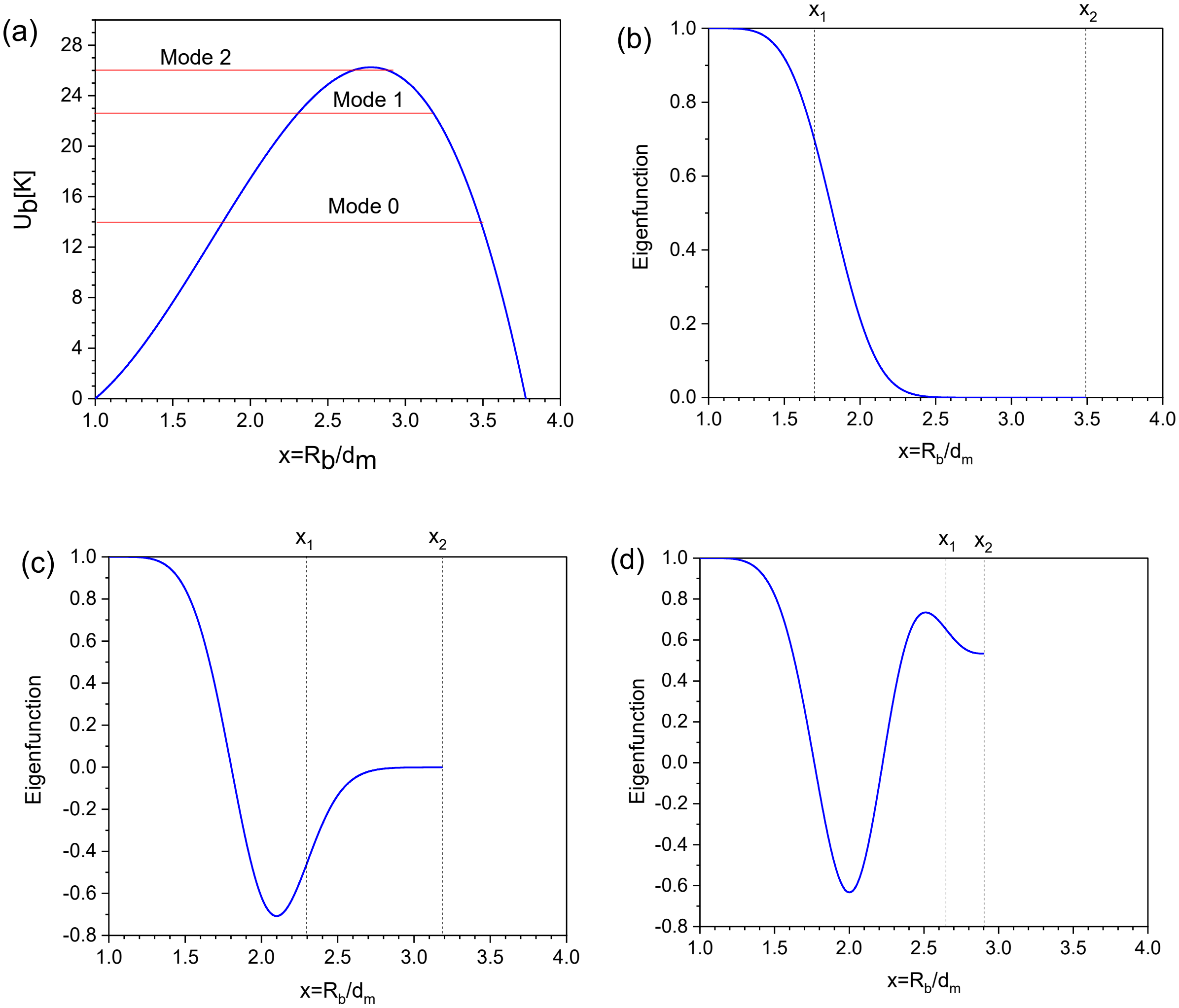

**Figure 5.** (a) The dependence of the potential $U_b$ on the coordinate $x = R_b/d_m$ for $^4$He at $P_- = 1.05$ MPa. The horizontal lines show the energies $E$ in degrees Kelvin (see equation (17) of mode 0, which corresponds to the zero state in (b); mode 1 corresponds to the first excited mode in (c); and mode 2 corresponds to the second excited mode (d). The vertical lines in figures (b), (c), and (d) correspond to the values of $x$ at which $U_b = E$. At $x > x_1$, the wave function decays. At $x > x_2$, the pore radius begins to grow, as the surface tension cannot compensate for the tensile forces associated with negative pressure.

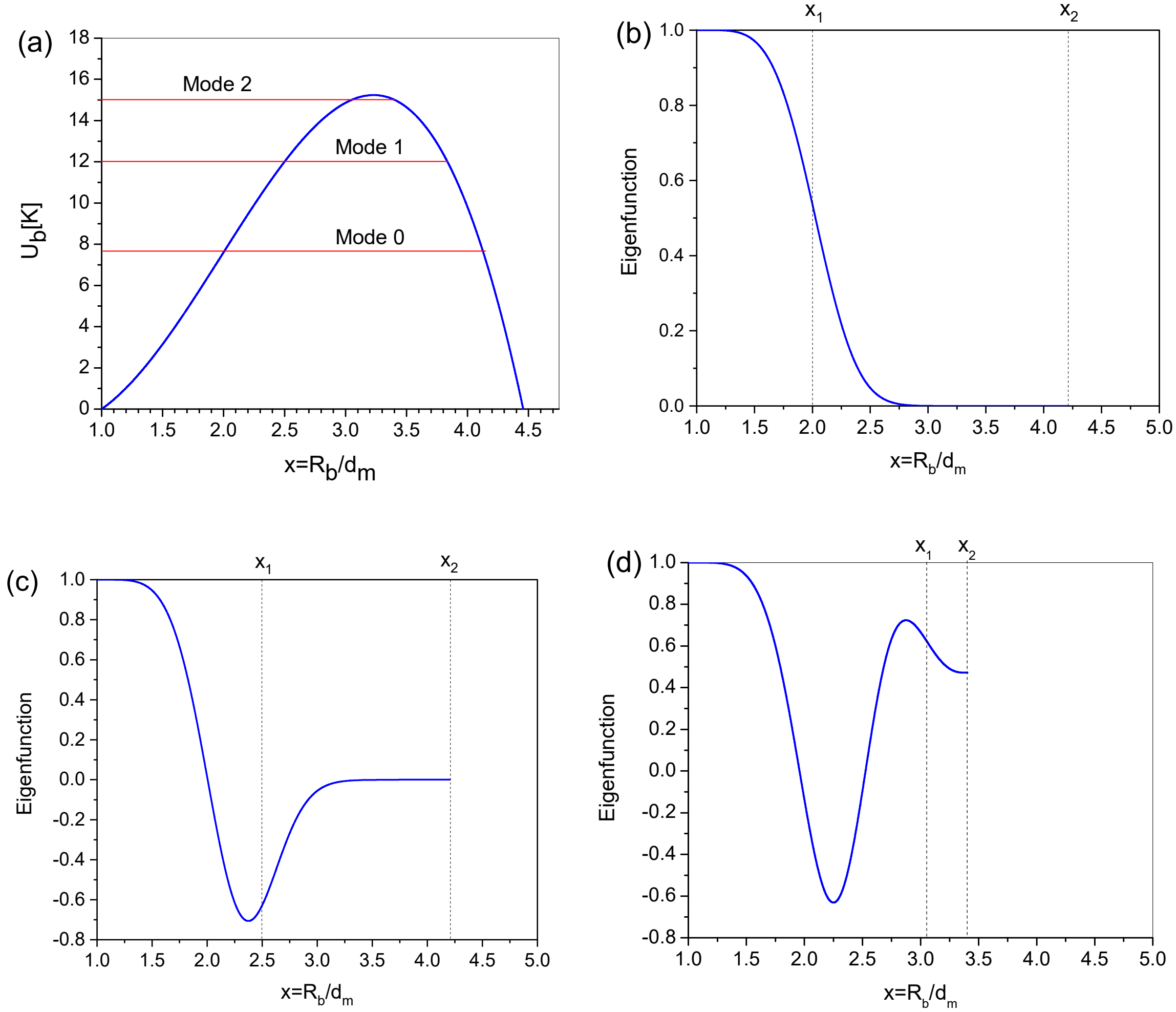

**Figure 6.** (a) The dependence of the potential $U_b$ on the coordinate $x = R_b/d_m$ for $^3$He at $|P_| = 0.4$ MPa. The horizontal lines show the energies $E$ in degrees Kelvin (see equation (17) of mode 0, which corresponds to the zero state in (b); mode 1 corresponds to the first excited mode in (c); and mode 2 corresponds to the second excited mode). The vertical lines in panels (b), (c), and (d) correspond to the values of $x$ at which $U_b = E$. At $x > x_1$, the wave function decays. At $x > x_2$, the pore radius begins to grow, as the surface tension cannot compensate for the tensile forces associated with negative pressure.

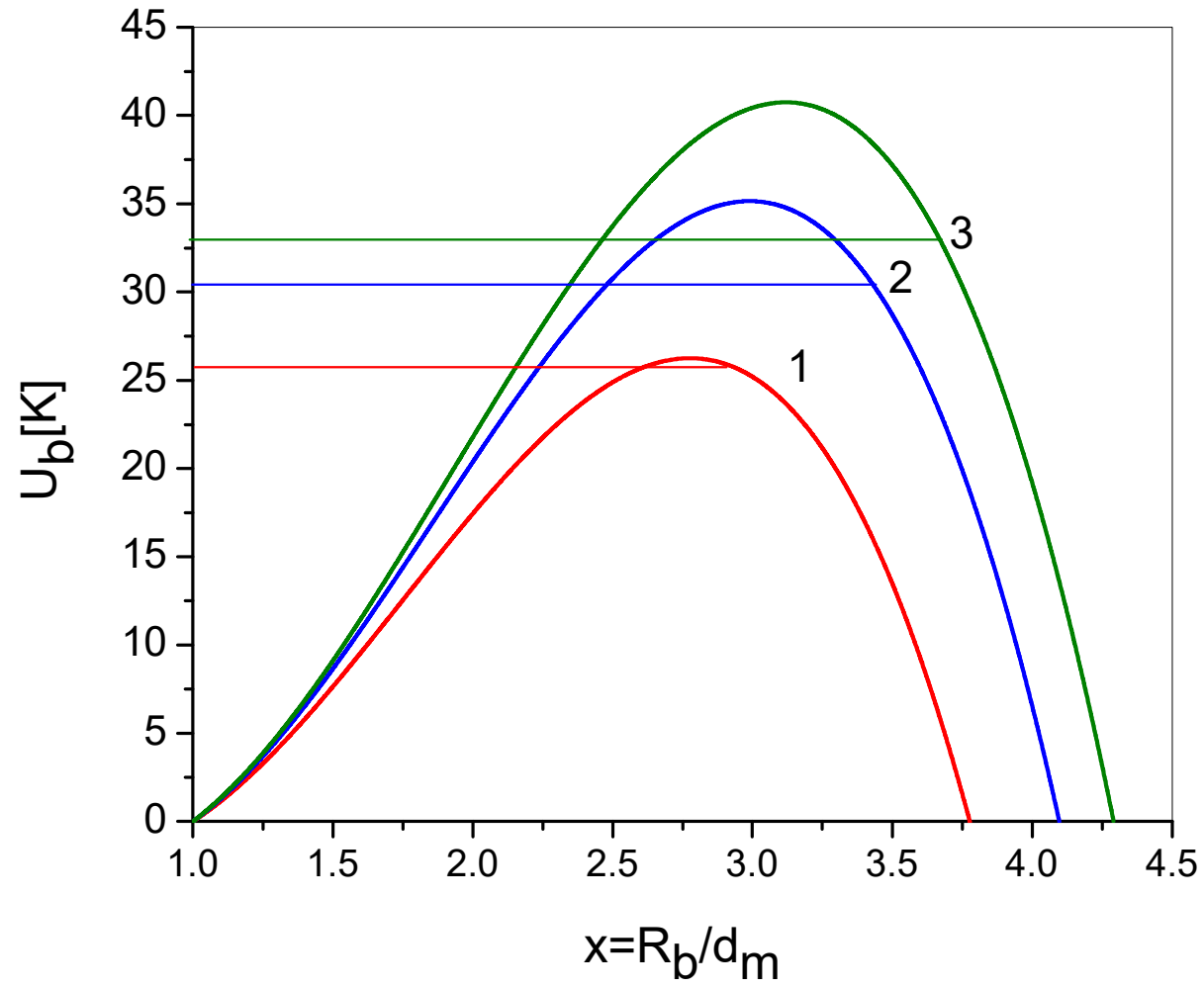

**Figure 7.** The dependence of the potential $U_b$ on the coordinate $x = R_b/d_m$ for $^4$He. 1, 2, and 3 correspond to mode 2 at pressures $|P_| = 1.05$, 1.04, and 1.02 MPa, respectively.

## IV. Nucleation probability

In a classical thermodynamic equilibrium ensemble of micropores, the probability of detecting a cavitation micropore with energy $E$ is determined by the Boltzmann distribution, $F(E) \propto \exp\left(-\frac{E}{T}\right)$, where $E$ is in degrees Kelvin. If the height of the barrier $U_0$ is much greater than the temperature $T$, then the probability that a particle will leave a potential well (i.e., overcome the barrier) is exponentially small. In chemistry, for example, the classical Arrhenius temperature dependence explains the sharp decrease in the rate of chemical reactions with decreasing temperature.

Now, let us turn to quantum mechanical considerations. Again, consider the "particle" in a potential well. In this well, the energy values of the particle are discrete. If $|P_| = 0$, then $\beta = 0$. In this case, as shown in Fig. 1(a), the particle cannot be outside the potential well. When $|P_|$ is not zero ($\beta > 0$), a particle has a non-zero probability of being outside the potential hole (Fig. 1(b)). Moreover, if its movement is limited by walls inside a potential well, then outside the well, it is not limited and can freely move along the $x$ −axis. In our case, a cavitation micropore (a hollow sphere with added mass) can increase in size unlimitedly.

According to [14–16], the number of cavitation pores formed per unit time in a unit volume, $n_k$, corresponding to the $k$-mode, is equal to:

$$\frac{dn_k}{dt} = \Gamma_k = k_B \frac{E_k}{\hbar} \frac{|\Psi_k(R_{*,k})|^2}{\frac{4\pi}{3}\int_0^{R_{*,k}} |\Psi_k|^2 r^2 dr} e^{-(E_k - E_0)/T} = k_B \frac{E_k}{\frac{4\pi}{3} R_{*,k}^3 \hbar} \frac{|\chi_k(R_{*,k})|^2}{\frac{1}{R_{*,k}} \int_0^{R_{*,k}} |\chi_k|^2 dr} e^{-(E_k - E_0)/T}, \tag{18}$$

where $R_{*,k}$ is the radius of the cavitation bubble corresponding to the $k$-mode. It is determined by the size of the bubble at which its growth begins (point $x_2$ in Figure 1). The total number of bubbles of critical size born per unit time per unit volume (nucleation rate) is correspondingly equal:

$$\Gamma = \sum_{k=1}^{K} \frac{dn_k}{dt} = \sum_{k=1}^{K} \Gamma_k. \tag{19}$$

Here, $K$ is the total number of modes that correspond to a given negative pressure value, $P_$.

Figure 8 shows the dependences of Γ on $|P_{-}|$ for $^{4}$He and $^{3}$He. Figures 7(a) and 7(b) show the steps involved in "crossing the highest-energy mode of the potential barrier." Thus, to the left of point A, there are two eigenmodes of the solution to equation (15); to the right of point A, there is only one. To the left of point B, there are three modes; to the right of point B, there are two. To the left of point C, there are four modes; to the right of point C, there are three. To the left of point D, there are five modes; to the right of point D, there are four, and so on. Clearly, the bubble generation rate (Γ) should sharply fall to the right of the "mode annihilation" points, since the next energy mode should make the main contribution to bubble generation. As noted above, at low temperatures, only the zero mode contributes. The others do not contribute due to the exponential multiplier, $e^{-(E_k-E_0)/T}$, in formula (18). The coincidence of lines 1–4 in Figures 8(a) and 7(b) at $|P_{-}| > 1.0672$ and 0.4061 MPa, respectively, is due to the fact that, at these pressures, the potential barriers are so low that only one mode exists in the potential well. At temperatures below 0.2 K, all modes except zero mode give no contribution to nucleation for $^{4}$He and below a temperature of 0.1 K for $^{3}$He.

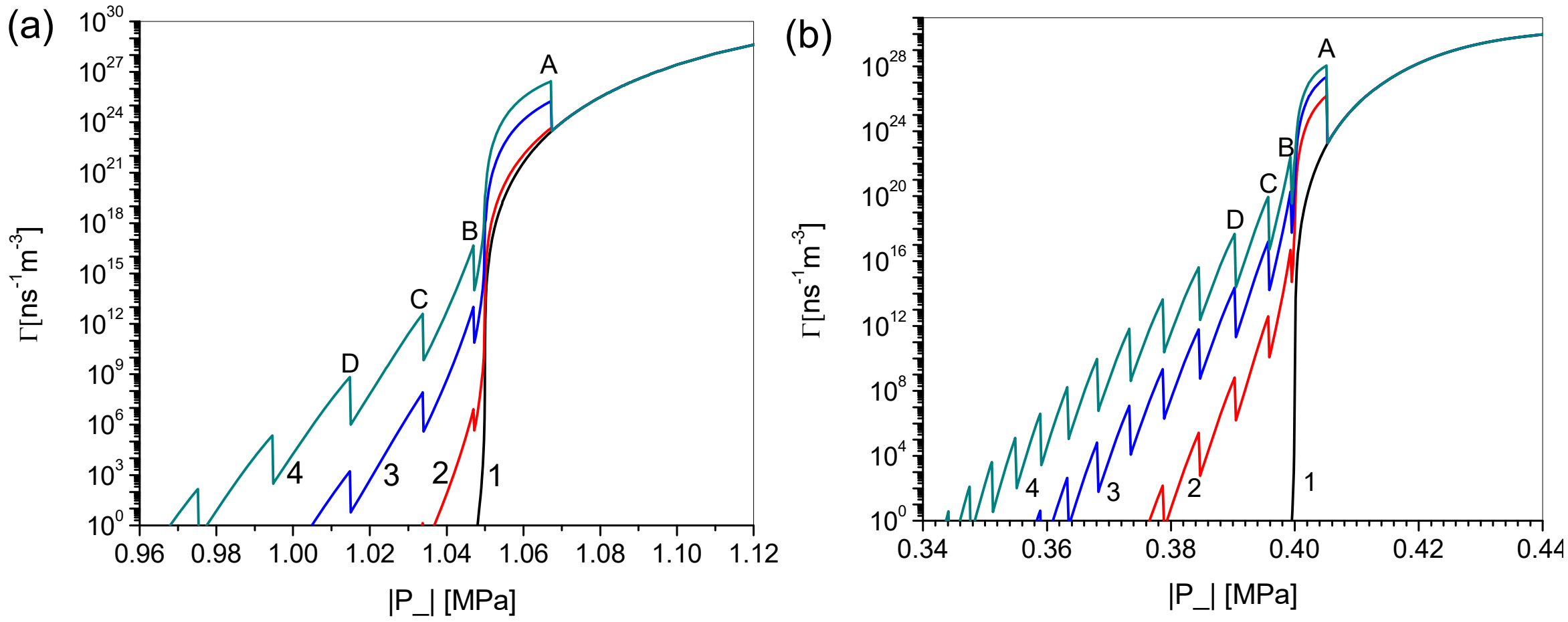

**Figure 8**. Dependence of Γ on $|P_{-}|$. (a) corresponds to $^{4}$He and (b) to $^{3}$He. Line 1 corresponds to $T = 0.001$ K, line 2 – $T = 0.3$ K, line 3 – $T = 0.4$ K, and line 4 – $T = 0.5$ K. To the left of point A, there are two eigenmodes of the solution to equation (17). To the right of point A, there is only one. To the left of point B, there are three modes, and to the right, there are two. To the left of point C, there are four modes, and to the right, there are three. To the left of point D, there are five modes, and to the right, there are four, and so on. The negative pressure at which jumps occur does not depend on temperature; only the magnitude of the jump depends on temperature.

Figure 9 shows the negative pressure region of dimension $l_{-}$, where cavitation can form. If the time of action of negative pressure in the region of cavitation generation is:

$$\tau_{-} < l_{-}/c_s, \tag{20}$$

then, the developing cavitation bubbles do not have time to displace the liquid from the generation region. Consequently, the pressure change becomes of the order of [14–16]

$$\Delta P \approx \Delta\rho c_s^2 \approx \rho c_s^2 \frac{\Delta V}{l_{-}^3}. \tag{21}$$

Here, $c_s$ is the speed of sound in the fluid.

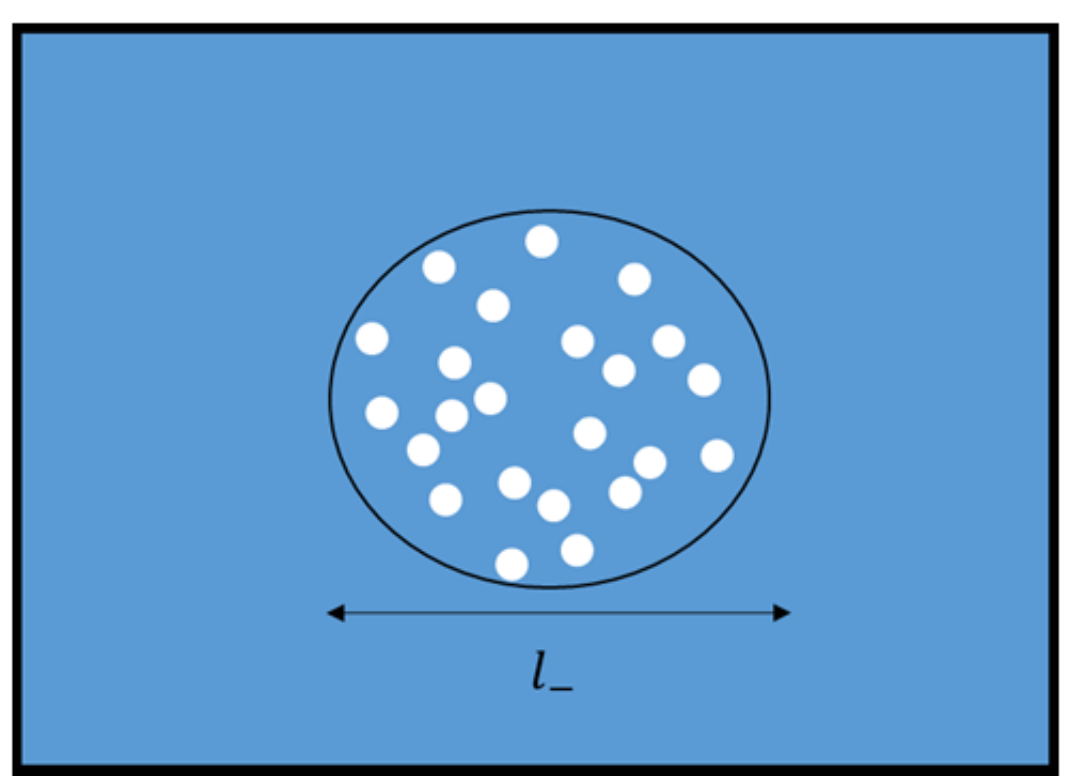

**Figure 9.** Schematic region of bubble generation where the absolute value of the negative pressure exceeds the critical pressure.

In accordance with the results shown in Figure 5(b)–(d), we can expect the initial radius of the incipient cavitation bubble to lie within the range of ~2.5–3.5 $d_m$. Without loss of generality, we will assume that, for all modes, the initial radius is $R_{b,0} = 2.5d_m$. Consequently, the volume of the emerging bubble is:

$$V_{b,0} = \frac{4}{3}\pi R_{b,0}^3 = \frac{4}{3}\pi\,(2.5d_m)^3 = 2.2\ \text{nm}^3. \tag{22}$$

We will assume that the distance between cavitation bubbles is much larger than their radius. In other words, all bubbles will expand independently of each other. In this case, the equation describing the change in bubble radius in the negative pressure region is [13]:

$$\frac{d}{dt}\left(\rho R_b^3\left(\frac{dR_b}{dt}\right)^2\right) = R_b^2\left(|P_-| - \frac{2\sigma_s}{R_b}\right)\frac{dR_b}{dt}. \tag{23}$$

Accordingly, the equation describing the pressure change in the negative pressure region is [21]:

$$\frac{d|P_-|}{dt} = -c_s^2\rho_{|P_-|}\left(V_{b,0}\,\Gamma + 4\pi\int_{\tau=0}^{t}\Gamma(\tau)\left(R_b(t-\tau)\right)^2\frac{dR_b(t-\tau)}{d(t-\tau)}\,d\tau\right). \tag{24}$$

Equations (17)–(19) and (22)–(24) completely describe the variation of negative pressure and bubble volume in the region, shown schematically in Figure 7. According to (24), the number of bubbles per unit volume and their total volume change over time as:

$$N_b = \int_0^t \Gamma(\tau)\,d\tau, \tag{25}$$

and

$$V = V_{b,0}\,\Gamma + 4\pi\int_{\tau=0}^{t}\Gamma(\tau)\left(R_b(t-\tau)\right)^2 v(t-\tau)d\tau. \tag{26}$$

Here, $v = \frac{dR_b}{dt}$ is the expansion velocity of the bubble.

Figure 10 shows the calculated results for $^4$He at a temperature of $T = 0.001$ K, which corresponds to curve 1 in Figure 6(a). It is assumed for certainty that the initial negative pressure is $P_- = -1.06$ MPa. As can be seen, the generation of cavitation bubbles stops after ~0.3 ns, and most of the bubbles have a radius of 50–60 nm.

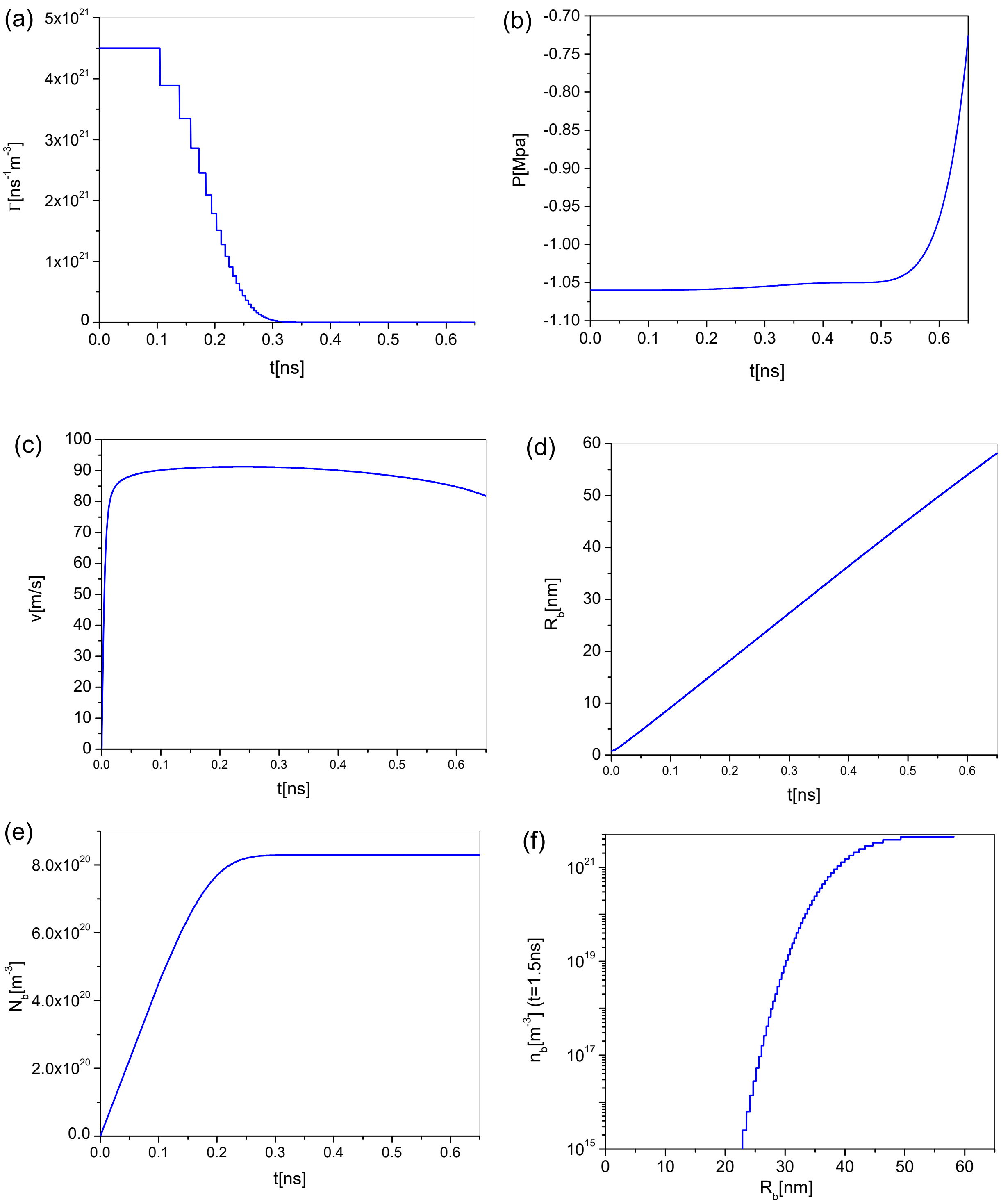

**Figure 10.** The calculation results are shown for $^4$He at $T = 0.001$ K (line 1 in Fig. 6(a)) and an initial negative pressure of $P_- = -1.06$ MPa. (a)–(e) show the time dependencies of the nucleation rate, negative pressure, bubble expansion rate, bubble radius, and total number of bubbles per unit volume, respectively. (f) shows the distribution of bubble sizes at time $t$ =0.65 ns.

Figure 11 shows the results of the $^4$He calculations at an initial negative pressure of $P_{_} = -1.06$ MPa, but at a higher temperature of $T = 0.5$ K. This corresponds to curve 2 in Figure 6(a). The jumps in bubble density in Figure 11(f) are due to the fact that the number of bubbles of a given radius is proportional to the nucleation rate $\Gamma$ at the time of their formation. Since there are jumps in $\Gamma$ in Figures 6(a) and 9(a) due to the appearance of new modes at pressure changes, $n_b$ also has them.

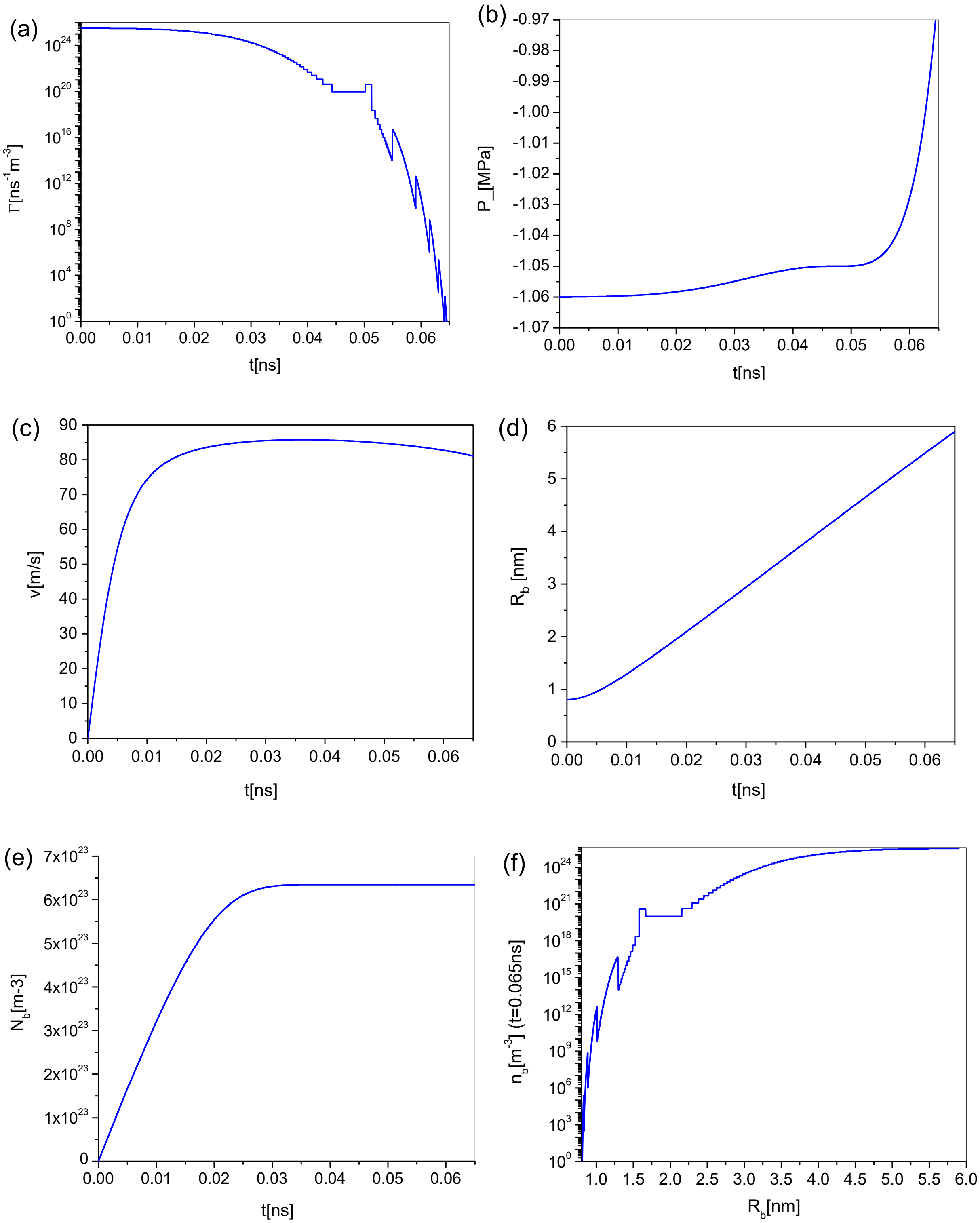

**Figure 11.** The calculation results are shown for [4]He at $T = 0.5$ K (line 2 in Fig. 6(a)) and an initial negative pressure of $P_{-} = -1.06$ MPa. (a)–(e) show the time dependencies of the nucleation rate, negative pressure, bubble expansion rate, bubble radius, and total number of bubbles per unit volume, respectively. (f) shows the distribution of bubble sizes at time $t = 0.65$ ns.

The temperature dependence of the critical pressure at which cavitation begins in [4]He and [3]He is given in [2]. The wide scatter in the experimental data can be explained by the presence of steps in the nucleation rate values Γ shown in Figures 6 and 9(a). However, according to our calculations, at very low temperatures when only the ground state is relevant, the scatter in the data should be much smaller.

A few words should be said about the difference between quantum and classical nucleation. As mentioned earlier, the rate of classical cavitation bubble formation is given by the equation: $\Gamma_{classic} \sim e^{-W_{cr}/k_B T}$, where $W_{cr} = 16\pi\sigma/3P_{-}^2$. Therefore, the rate of classical cavitation bubble formation, or $\Gamma_{classic}$, is a continuous function of negative pressure. As shown in Fig. 8, quantum cavitation is accompanied by jumps within a certain temperature range. At an increase in the absolute value of negative pressure $|P_{-}|$, the rate of bubble formation Γ continuously increases and then sharply decreases. This is due to the finite number of modes in the potential well at non-zero $|P_{-}|$. At very low temperatures, Γ becomes a continuous function of negative pressure again because, in this case, only the zero mode contributes to nucleation.

The tunnel cavitation mechanism dominates at cryogenic temperatures. As we said earlier, as the temperature increases, the nucleation of cavitation follows the Arrhenius law. Now, let us estimate the liquid helium temperatures at which the quantum nucleation mechanism is replaced by the classical one. According to equation (18), the probability of forming a cavitation rupture from the *k*-th level is proportional to the product of the probability of filling this level, $w_{T,k} \sim e^{-(E_k - E_0)/T}$ multiplied by the probability of a tunneling transition from this level, $w_{tun,k} \sim \frac{|\chi_k(R_{*,k})|^2}{\frac{1}{R_{*,k}}\int_0^{R_{*,k}} |\chi_k|^2 dr}$:

$$w_k \sim w_{T,k} \cdot w_{tun,k}. \qquad (27)$$

Thus, in this case, the probability of cavitation pore formation from the zero level is determined solely by tunneling

$$w_0 \sim \frac{|\chi_0(R_{*,0})|^2}{\frac{1}{R_{*,0}}\int_0^{R_{*,0}} |\chi_0|^2 dr}. \qquad (28)$$

When the energy of the *k*-mode is close to the value of the potential barrier, $E_k \approx U_b$, the probability of a sub-barrier transition is close to unity, and, therefore,

$$w_k = w_{U_b} \sim e^{-(U_b - E_0)/T}. \qquad (30)$$

In this case, we can consider $w_{U_b}$ to be analogous to the Arrhenius probability for chemical reactions. Obviously, $U_b$ cannot be less than the energy of the first mode $E_1$. By comparing $w_{U_b}$ with $w_0$, we can determine whether the Arrhenius law is dominant to cavitation initiation:

$$e^{-(U_b - E_0(U_b))/T} \sim w_0(U_b), \qquad U_b > E_1(U_b). \qquad (31)$$

Estimates show that the transition from tunneling to the Arrhenius law occurs at $T > 0.1$ K for $^3$He and at $T > 0.4$ K for $^4$He.

It should be noted that the Schrödinger equation was also used in [22] to describe cavitation nucleation. However, they did not account for the dependence of surface tension on bubble radius and pressure. Unfortunately, [22] does not provide the energy values corresponding to the different eigenwave functions or the wave functions themselves. This makes it difficult to compare our results with those obtained in [22].

## Conclusions

In this paper, based on the Schrödinger-like equation, we consider the tunneling mechanism of cavitation formation in liquid helium and obtain threshold values of negative pressure as a function of temperature for $^3$He and $^4$He. We present the results of calculating the surface tension coefficients for flat and curved interfaces in the Lennard-Jones interaction potential approximation. We demonstrate that the temperature dependence of the critical pressure (the pressure at which cavitation begins) is stepwise. Calculations that take into account the curvilinear boundary and stretching of the liquid in a negative pressure field are in satisfactory agreement with experimental data [2].

## References

[1] Y. B. Zel'dovich, Theory of formation of a new phase. Cavitation, Zh. Eksp. Teor. Fiz. **12**, 525, (1942)
[2] E. Caupin, S. Balibar, Cavitation pressure in liquid helium, Phys. Rev. B, **64**, 064507 (2001)
[3] V.I. Goldanski, Chemical reaction at very low temperature, Ann. Rev. Phys. Chern. **27**, 85 (1976)
[4] V.A. Benderskii, V.I. Goldanski, D.E. Markov, Quantum dynamics in low-temperature chemistry, Physics Report **233**, 195 (1993)
[5] F. Hund, Zur deutung der molekelspektren III, Z Phys. **43**, 805 (1927)
[6] I.M. Lifshitz, Yu. Kagan, Quantum kinetics of phase transitions at temperatures close to absolute zero, Sov. Phys. JETP, **35**, 206 (1972)
[7] J. Casulleras, J. Boronat, Progress in monte carlo calculations of fermi systems: normal liquid $^3$He, Phys. Rev. Lett. **84**, 3 (2000)
[8] G.H. Bauer, D.M. Ceperley, N. Goldenfeld, Path-integral Monte Carlo simulation of helium at negative pressures *Phys. Rev.* B **61** 9055 (2000)
[9] H.J. Maris, Theory of quantum nucleation of bubbles in liquid helium *J. Low Temp. Phys.,* **98**, 403 (1995)
[10] C.E. Campbell, R. Folk, E. Krotseheck, Critical behavior of liquid 4He at negative pressures, J. Low Temp. Phys. **105**, 13 (1996)
[11] M. Guilleumas, M. Barranco, D.M. Jezek, R.J. Lombard, M. Pi, Quantum cavitation in liquid helium *Phys. Rev.* B, **54,** 16135 (1996)
[12] K.R. Atkins, Y. Narahara Surface tension of liquid $^4$He, Phys. Rev, **138** 437 (1965).
[13] O. M. F. R. S. Lord Rayleigh, On the pressure developed in a liquid during the collapse of a spherical cavity, Philos. Mag. 34, 94 (1917)
[14] M.N. Shneider and M. Pekker, Surface tension of small bubbles and droplets and the cavitation threshold, arxiv.org/pdf/1901.04329
[15] M.N. Shneider and M. Pekker, Liquid Dielectrics in an Inhomogeneous Pulsed Electric Field. Dynamics, cavitation and related phenomena, 2nd Edition (IOP Publishing, Temple Circus, Temple Way, Bristol, UK, 2020)
[16] J.E. Lennard-Jones, On the determination of molecular fields: II: From the equation of state of a gas, Proc. Roy. Soc., **A106**, 463 (1924)
[17] S.W. Van Sciver, Helium Cryogenics (Springer, New York, Dordrecht, Heidelberg, London, 1986)

[18] W.E. Massey, Ground state of liquid helium – boson solution for mass 3 and 4, Phys. Rev. **151**, 153 (1966)
[19] D. Schiff, V. Verlet, Ground state of liquid Helium-4 and Helium-3, Phys. Rev. **160**, 208 (1967)
[20] K. Atkins, Liquid Helium (University Press, Cambridge, England, 1959)
[21] X. Zhang, J. Li, M.N. Shneider, Dynamics of homogeneous cavitation with pressure feedback, Phys. Fluids **34**, 101704 (2022)
[22] T. Nakamura, Y. Kanno, and S. Takagi, Single-collective-degree-of-freedom models of macroscopic quantum nucleation, Phys. Rev. B **51**, 8446 (1995)